\documentclass[final,3p,times,twocolumn,compress]{elsarticle}

\usepackage{amsmath,amssymb,amsbsy,amstext,amsfonts,mathrsfs,latexsym,bm,physics,graphicx}

\usepackage{color}
\definecolor{dullpurple}{rgb}{0.431,0.188,0.534}
\definecolor{dullred}{rgb}{0.706,0.208,0.192}
\definecolor{dullblue}{rgb}{0,0.298,0.49}
\definecolor{blue3}{RGB}{31,119,180}
\definecolor{red3}{RGB}{214,39,40}
\definecolor{green3}{RGB}{44,160,44}
\usepackage[colorlinks]{hyperref}
\AtBeginDocument{%
  \hypersetup{
    citecolor=dullblue,
    anchorcolor=blue3,
    filecolor=green3,
    menucolor=red3,
    linkcolor=dullred,   
    urlcolor=dullpurple}}

\makeatletter
\def\ps@pprintTitle{%
  \let\@oddhead\@empty
  \let\@evenhead\@empty
  \def\@oddfoot{\reset@font\hfil\thepage\hfil}
  \let\@evenfoot\@oddfoot
}
\makeatother

\begin{document}

\begin{frontmatter}

\title{A {\small{small}} Universe}

\author[a]{Jean-Luc Lehners}
\affiliation[a]{organization={Max Planck Institute for Gravitational Physics (Albert Einstein Institute)},
            city={Potsdam},
            postcode={D-14476},
            country={Germany}}

\author[b,c]{Jerome Quintin}
\affiliation[b]{organization={Department of Applied Mathematics and Waterloo Centre for Astrophysics, University of Waterloo},
            city={Waterloo},
            postcode={N2L 3G1},
            state={Ontario},
            country={Canada}}
\affiliation[c]{organization={Perimeter Institute for Theoretical Physics},
            city={Waterloo},
            postcode={N2L 2Y5},
            state={Ontario},
            country={Canada}}

\begin{abstract}
Many cosmological models assume or imply that the total size of the universe is very large, perhaps even infinite. Here we argue instead that the universe might be comparatively small, in fact not much larger than the currently observed size. A concrete implementation of this idea is provided by the no-boundary proposal, in combination with a plateau-shaped inflationary potential. In this model, opposing effects of the weighting of the wave function and of the criterion of allowability of the geometries conspire to favour small universes. We point out that a small size of the universe also fits well with swampland conjectures, and we comment on the relation with the dark dimension scenario.
\end{abstract}

\begin{keyword}
No-boundary proposal \sep Cosmic inflation \sep Quantum cosmology
\end{keyword}

\end{frontmatter}

\section{Introduction}
\label{introduction}

Perhaps the most basic characteristic of our universe is that it is big. When extrapolated to present times, the part of the universe that we see in the cosmic microwave background (CMB) has a size of a little over three times the current Hubble radius $c/H_0$ \cite{Gott:2003pf} (about $46$ billion light years), which is half a million times the size of the Milky Way. However, we do not know how large the universe actually is, as we cannot see beyond the limit imposed on us by the CMB. It has been suggested repeatedly that the universe might have a non-trivial topology, which would lead to specific patterns in the CMB if the shortest non-trivial loop fits within the observed last scattering surface ({\it e.g.}~\cite{Lachieze-Rey:1995qrb,Cornish:1997ab,Starkman:1998qx,Levin:2001fg}). Such patterns have been searched for, and no evidence of their existence has been found ({\it e.g.}~\cite{Cornish:2003db,ShapiroKey:2006hm,Bielewicz:2010bh,Vaudrevange:2012da,Aurich:2013fwa,Luminet:2013ama,Planck:2013okc,Planck:2015gmu,Luminet:2016bqv}). More subtle correlations might be detectable even if the non-trivial topological feature is larger than the region seen in the CMB, see \cite{COMPACT:2022gbl,COMPACT:2022nsu,COMPACT:2023rkp}. In the meantime, we will assume here that the universe possesses a simple topology (see also \cite{Galloway:2020eju}).

In the absence of observational evidence, we may ask: how large do we expect the universe to be, based on our current understanding of theoretical cosmology? This is the question that concerns us in the present paper. Needless to say, the word \emph{current} is crucial -- cosmology has seen spectacular progress over the last century, accompanied by repeated, significant changes in our outlook on the universe. Also, few complete cosmological models exist. Most models describe only what happens to a certain (comoving) region of the universe, without implications for (or restrictions on) how many such regions might exist in total. But if we want to address the question of the size of the universe, we must do so in a complete cosmological model.

A heuristic view that has been advocated repeatedly is that the universe might have started out at the Planck (energy) scale, with wild fluctuations in geometry and matter configurations. Somewhere, it is argued, the conditions will turn out to be right for inflation to start and the corresponding region of the universe then develops into the large universe we see. However, it is typically not said how this beginning is supposed to have occurred and how many favourable regions one might expect to develop. In some sense, eternal inflation improved upon this scenario by showing that, if it occurs, the late time characteristics of the universe will essentially be independent of the beginning, as we would then be removed from the beginning by a possibly infinite number of tunneling events, in which a new inflating spacetime region of the universe is created from a preexisting inflationary patch of the universe. In this kind of model, the universe is then predicted to be spatially infinite.

This infinity leads to paradoxes, for example with regard to assigning probabilities in quantum theory, as in an infinite universe events that are not forbidden occur infinitely often \cite{Guth:2007ng,Page:2009qe}. In fact the general question of a measure on such a set of pocket universes remains completely unresolved and may well be misguided. Specifying a measure assumes the existence of an overall slicing, which may be seen as an overall time coordinate in this multiverse. If one imagines one, then regions with the same field configurations exist at various `times', and in fact a given set of field configurations is reproduced arbitrarily often globally. This causes the action to diverge \cite{Jonas:2021xkx}, which signals a semi-classical inconsistency of the model. For these reasons, we deem eternal inflation to be physically inconsistent.

Another complete cosmological model that has been proposed is that of a cyclic universe, in which the universe is rendered smooth during ekpyrotic contracting phases and in which local quantities evolve cyclically \cite{Steinhardt:2001st}. This model also assumes a spatially infinite universe. Because the smoothing occurs at a low energy scale it avoids the worst issues with assigning a measure \cite{Johnson:2011aa}, though the problem with assigning quantum probabilities remains. For this kind of model, the action is also found to diverge \cite{Jonas:2021xkx}, which we again take as a sign of semi-classical inconsistency. Moreover, the physics of the bounce phase remains insufficiently understood.

\begin{figure}[ht]
	\centering
	\includegraphics[width=0.45\textwidth]{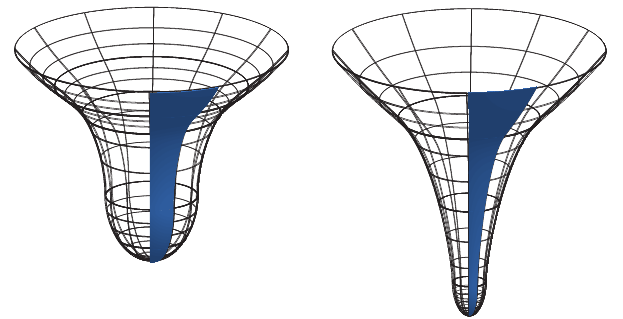}
	\caption{The current size of the universe is the same for models with different inflationary/nucleation scales, as long as the amount of inflation is minimal. In this cartoon, the spatial part of the universe is a circle rather than a $3$-sphere. On the left we have a model with a low inflationary energy scale and on the right a higher scale. The observable part of the universe is shaded in blue.}
	\label{fig:cartoon}
\end{figure}

This leaves us with models of ordinary, non-eternal inflation, which however must be completed in the sense that their beginning must be explained (it is not possible to assume that ordinary inflation was the first phase of evolution of the universe because this leads both to a classical singularity \cite{Borde:2001} -- see \cite{Yoshida:2018ndv,Lesnefsky:2022fen,Geshnizjani:2023hyd} for recent developments -- and a semi-classical instability \cite{DiTucci:2019xcr}). This brings us to the best understood theories of the origin of the universe: the no-boundary \cite{Hartle:1983ai} and tunnelling proposals \cite{Vilenkin:1983xq}. In both of these frameworks spacetime and matter emerge from a quantum tunnelling event out of nothing. The big bang singularity is replaced by a smooth, regular, quantum geometry which includes a Euclidean (or near-Euclidean) region where the big bang would have been. For a cartoon of the idea see Fig.~\ref{fig:cartoon}, and for a detailed description see \cite{Lehners:2023yrj}. Because these no-boundary field configurations are both compact and regular, the action is automatically finite. Nevertheless, in the case of the tunnelling proposal the wave function favours anisotropic configurations to such an extent that the wave function becomes non-normalisable, and for this reason we will focus on the no-boundary wave function here, for which isotropic universes receive the highest weighting. In this sense a no-boundary origin immediately explains the observed spatial isotropy on the largest scales.

An intriguing feature of the no-boundary wave function is that it leads to the appearance of classical, Lorentzian universes only in the presence of a dynamical attractor \cite{Hartle:2008ng}. Inflation can play precisely this role. (Ekpyrotic cosmology could too \cite{Battarra:2014xoa,Battarra:2014kga,Lehners:2015efa}.) In fact, only a few $e$-folds of inflationary evolution are required to localise the wave function on Lorentzian histories. However, in order to explain the present-day features of our universe, in particular its spatial flatness, a longer inflationary phase is required. This is because the basic no-boundary requirements of compactness and regularity force the universe to have closed spatial sections (as reviewed in, {\it e.g.}, \cite{Lehners:2023yrj}). In other words, given our assumption of a trivial spatial topology, on large scales the universe is described by a $K=+1$ Robertson-Walker metric with spatial $3$-spheres\footnote{Other topologies with closed spatial sections are possible. An example is $S^2\times S^1,$ which has been shown to be suppressed compared to $S^3$ in the no-boundary wave function \cite{Halliwell:1990tu,Bousso:1995cc}. We should note that the contribution of more intricate topologies has not been widely explored to date.},
\begin{equation}
    \mathrm{d}s^2 = -\mathrm{d}t^2 + a(t)^2 \mathrm{d}\Omega_{(3)}^2\,.
\end{equation}
This feature may be at odds with observations, which show that the average curvature is currently very small. It is useful to look at these aspects more quantitatively now.

\section{Size of the Universe}

The evolution of the universe may be described by the Friedmann equation
\begin{equation}
	1=\Omega_K(a)+\Omega_\phi(a)+\Omega_\mathrm{r}(a)+\Omega_\mathrm{m}(a)\,,
\end{equation}
with
\begin{align}
	\Omega_K&=-\frac{1}{a^2H^2}\,,\qquad\Omega_\phi=\frac{1}{3H^2}\left(\frac{1}{2}\dot\phi^2 + V(\phi) \right)\,,\nonumber\\
	\Omega_\mathrm{r}&=\frac{\rho_\mathrm{r,0}}{3a^4H^2}\,,\qquad\,\Omega_\mathrm{m}=\frac{\rho_\mathrm{m,0}}{3a^3H^2}\,.
\end{align}
Here a dot denotes a derivative with respect to the time coordinate $t,$ $H\equiv\dot{a}/a$, and we have assumed the presence of a scalar field $\phi$ with a potential $V(\phi)$. We have also included pressure-free matter and radiation, with present-day densities $\rho_{\mathrm{m},0}$ and $\rho_{\mathrm{r},0}$ respectively, which in these models are created at reheating. We are ignoring dark energy, though, since it is negligible for the present discussion. Finally, we are working in units where Newton's gravitational constant $G$ is set to $1/(8\pi)$ and where the speed of light is set to unity.

Observations show that the curvature term is currently very small, the {\it Planck} CMB data alone\footnote{There has been stronger evidence for $\Omega_{K,0}<0$ from CMB data alone \cite{Planck:2018vyg}, but when combined with baryonic acoustic oscillations data one recovers consistency with a flat universe at the cost of tension between the datasets -- for different views see, {\it e.g.}, \cite{Park:2017xbl,DiValentino:2019qzk,Handley:2019tkm,Efstathiou:2020wem,Vagnozzi:2020dfn,Vagnozzi:2020rcz}.}
finding \cite{Planck:2018jri}
\begin{equation}
    \Omega_{K,0} = -0.011_{-0.012}^{+0.013} \quad (95\% \, \textrm{CL})\,.
\end{equation}
This is puzzling, as the curvature term has been growing during the phases of radiation ($\propto a^2$) and matter ($\propto a$) domination. A standard resolution of this puzzle is to invoke an inflationary phase, during which the curvature term shrinks. For a slow-roll no-boundary history the Lorentzian evolution coincides pretty closely with that of a closed de Sitter universe at constant vacuum energy $V(\phi_\mathrm{infl})$, with scale factor
\begin{equation}
    a(t)=\frac{1}{H_\mathrm{infl}}\cosh\left(H_\mathrm{infl}t\right)\,,
\end{equation}
where the nucleation scale (which to a first approximation can be thought of as the onset of the classical evolution) is given by
\begin{equation}
    a_\mathrm{nucl}\equiv a(t=0)=\frac{1}{H_\mathrm{infl}}=\sqrt{\frac{3}{V(\phi_\mathrm{infl})}}\,.
\end{equation}
Thus, during the inflationary phase the curvature component is given by
\begin{equation}
    \Omega_K=-\csch^2\left(H_\mathrm{infl}t\right) \approx -e^{-2N}\,,
\end{equation}
where we defined the number of $e$-folds of inflation as $N\equiv\ln(a/a_\mathrm{nucl})$, and where the approximation holds at late times. Importantly, we see that the curvature scales as an exponential of the $e$-folding number during inflation.

Putting the various phases together, we find that at present times we may expect the curvature term to be given by
\begin{align}
    \Omega_{K,0}&\approx -e^{-2N_\mathrm{infl}}\left(\frac{a_\mathrm{eq}}{a_\mathrm{reh}}\right)^2\left(\frac{a_0}{a_\mathrm{eq}}\right)\nonumber\\
    &=-e^{-2N_\mathrm{infl}}\left(\frac{T_\mathrm{reh}}{T_\mathrm{eq}}\right)^2(1+z_\mathrm{eq})\,,
\end{align}
where the duration of inflation from nucleation to reheating is $N_\mathrm{infl}=\ln(a_\mathrm{reh}/a_\mathrm{nucl})$. Here $T_\mathrm{eq}$ and $z_\mathrm{eq}$ denote the temperature and redshift at matter-radiation equality. As one can see, the value of the curvature term depends crucially on the amount of inflation. In fact, we can turn this relation around and find the minimal number of $e$-folds of inflation $N_\mathrm{infl,min}$ required to fit the current observational bounds on the curvature. Using $1+z_\mathrm{eq}\approx 3400$, $T_\mathrm{eq}\approx 0.8\,\mathrm{eV}$, one finds (see also \cite{WMAP:2008lyn,Planck:2018jri})
\begin{align}\label{eq:Ninflmin}
    N_\mathrm{infl,min} & \approx 34 + \ln\left(\frac{T_\mathrm{reh}}{1\,\mathrm{TeV}}\right) - \frac{1}{2}\ln\left(\frac{\Omega_{K,0}}{-0.01}\right) \nonumber \\
    & \approx 62 + \ln\left(\frac{T_\mathrm{reh}}{10^{15}\, \textrm{GeV}}\right) - \frac{1}{2}\ln\left(\frac{\Omega_{K,0}}{-0.01}\right)\,.
\end{align}
This minimal amount depends sensitively on the scale of inflation, which we have assumed coincides with the scale of reheating.

What is the implied current size of the universe in these models? To determine this, we start from the size at nucleation and factor in the subsequent evolutionary phases:
\begin{equation}
    a_0 = a_\mathrm{nucl} \, e^{N_\mathrm{infl}}\, \left(\frac{T_\mathrm{reh}}{T_\mathrm{eq}}\right) (1+z_\mathrm{eq})\,.
\end{equation}
Now $a_\mathrm{nucl}=H_\mathrm{infl}^{-1}\sim T_\mathrm{reh}^{-2}$. If we assume for the moment that there was a minimal amount of inflation [{\it cf.}~\eqref{eq:Ninflmin}], then using $\exp(N_\mathrm{infl,min}) \propto T_\mathrm{reh}$ we see that the current size of the universe becomes independent of the nucleation scale, {\it i.e.}~independent of the scale of inflation. This happens because a higher inflationary scale implies a smaller nucleation size but also a more pronounced inflationary expansion, and the two effects exactly compensate each other -- see Fig.~\ref{fig:cartoon} for an illustration. In fact, by assumption, the size comes out as being on the order of the currently observable scale,
\begin{equation}
    a_0=\frac{1}{H_0\sqrt{-\Omega_{K,0}}}\,.
\end{equation}
Thus, if inflation lasted not much longer than necessary to resolve the flatness puzzle, the size of the universe is just a few times larger than the currently observable size, with a closed spatial geometry.

\section{Allowability}

But the crucial question is: why would we expect a small number of $e$-folds of inflation? In the no-boundary framework, inflationary histories obtain a probability $P(\phi_\mathrm{i})$ that depends on the location of the scalar field in the inflationary potential at the onset of the inflationary phase, according to
\begin{equation}
    P(\phi_\mathrm{i}) \propto \exp(\frac{24\pi^2}{\hbar V(\phi_\mathrm{i})})\,.
\end{equation}
(Here $\phi_\mathrm{i}$ is the real part of the scalar field at the South Pole of the instanton.) This expression shows that low values of the potential come out as vastly preferred. Typically, this will imply an inflationary phase that is much too short to resolve the flatness puzzle and to generate appropriate cosmological perturbations. However, it has recently been realised that not all no-boundary instantons are likely to make sense in quantum gravity, as some of them develop an instability \cite{Lehners:2022xds,Hertog:2023vot}.

No-boundary instantons are generally described by complex instantons. In the context of quantum field theory, Kontsevich \&~Segal \cite{Kontsevich:2021dmb} proposed that complex metrics $g_{\alpha\beta}$ should only be allowed if they lead to convergent path integrals once all possible ($p$-form) matter types have been included. Explicitly, their criterion reads
\begin{align}
    & \Big|\exp\Big(\frac{i}{\hbar}S\Big)\Big|<1  \,\,\, \textrm{implying} \nonumber \\
    & \Re\left(\sqrt{g} \, g^{\alpha_1 \beta_1} \cdots g^{\alpha_{p+1} \beta_{p+1}} F_{\alpha_1 \cdots \alpha_{p+1}} F_{\beta_1 \cdots \beta_{p+1}}\right) > 0\,, \label{KS}
\end{align}
where $S$ is the action and $F$ is the $(p+1)$-form field strength in $D$ dimensions. Pointwise, one can write the metric in diagonal form $g_{\alpha\beta} = \delta_{\alpha\beta} \lambda_\alpha$ (no summation), and then the above criterion simplifies to the condition
\begin{equation}
    \Sigma \equiv \sum_{\alpha=1}^D \left|\mathrm{Arg}(\lambda_{\alpha})\right| < \pi\,. \label{bound}
\end{equation} 
Roughly speaking, one can say that the metric is allowed to be complex, but not too much.

\begin{figure}[ht]
	\centering
	\includegraphics[width=0.4\textwidth]{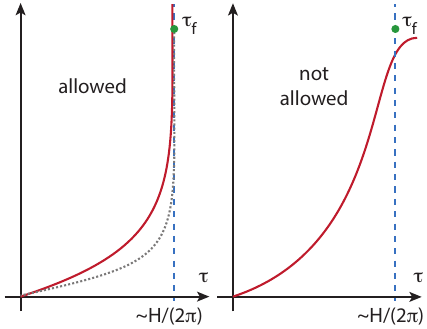}
	\caption{This figure shows the complex time plane with $\Re(\tau)$ being Euclidean time and $\Im(\tau)$ Lorentzian time. The origin corresponds to the South Pole of a no-boundary solution, and $\tau_\mathrm{f}$ shows the location of the final time (this can be thought of as the spatial section at the end of inflation). In red we draw the maximal $\Sigma=\pi$ curve. If the final time $\tau_\mathrm{f}$ can be reached from the origin while remaining below the red curve at all times, then the solution is allowed (left panel, the dotted line shows a possible allowed path), while otherwise it is not allowed (right panel). The blue dashed line indicates the Euclidean time location of the final point.}
	\label{fig:path}
\end{figure}

If one extends this criterion to quantum gravity, as Witten \cite{Witten:2021nzp} has suggested, then one finds that certain types of metrics are not allowed, precisely because matter path integrals would diverge on such backgrounds. Concretely, one can determine the allowability as follows (for details, see \cite{Jonas:2022uqb,Hertog:2023vot}). For the case of interest to us, the metric depends only on (complexified) time. A particular solution may then be represented by a path in the complex time plane -- the details of this calculation can be found in \ref{app:details}. In the no-boundary case this is a path from the origin (where we choose the South Pole to be located) up to the final point, which represents the late-time hypersurface on which the evolution ends; see Fig.~\ref{fig:path}. One should now picture two curves, emanating from the origin and saturating the inequality \eqref{bound}, {\it i.e.}~two lines with $\Sigma=-\pi$ and $\Sigma=+\pi.$ The actual solution is allowable if it lies in between these two extremal lines. It turns out that in the case of interest to us only the maximal line ($\Sigma=+\pi$) is important, the other extremal line lying at $\Im(\tau)<0$. If the actual solution lies entirely below that line, the solution is allowed.

This has consequences for no-boundary solutions. As was shown in \cite{Witten:2021nzp,Lehners:2021mah}, pure de Sitter instantons are always allowed. However, as discussed in \cite{Lehners:2022xds,Hertog:2023vot}, this is no longer true when a scalar field with a potential is added. More specifically, in \cite{Hertog:2023vot} it was shown that only certain types of potentials (mostly concave potentials) admit allowable no-boundary solutions that lead to $60$ $e$-folds of inflationary expansion. The preference for concave/plateau-type potentials may be understood as follows: when the potential is steeper, it requires a more complex scalar field. For instance, at the `bottom' (South Pole) of the no-boundary instanton, to a first approximation the scalar field value is given by
\begin{equation}
    \phi_\mathrm{SP} \approx \phi_\mathrm{i} - \frac{\pi}{2}\frac{V_{,\phi}}{V}\, i = \phi_\mathrm{i} - \frac{\pi}{2}\sqrt{2\epsilon}\, i\,,\label{eq:phiSP}
\end{equation}
where $\epsilon \equiv (V_{,\phi})^2/(2V^2)$ is the slow-roll parameter. A more complex scalar implies a more complex metric in turn, and this fact then disallows no-boundary instantons starting out on a steep slope of the potential.

In the present paper we do not take the amount of inflation for granted. Rather, we ask: for a given potential, how many $e$-folds of inflation are required in order for the associated no-boundary instantons to be allowed? When the potential is of plateau type, then a longer inflationary phase means starting out on a flatter region of the potential and hence will improve the allowability properties of the associated no-boundary solutions.

\section{Example}

\begin{figure}[ht]
	\centering
	\includegraphics[width=0.38\textwidth]{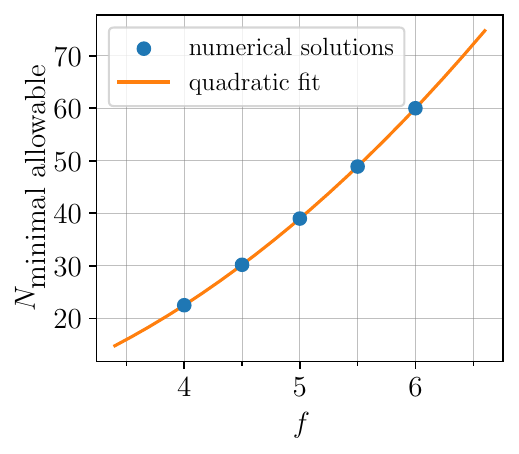}
	\caption{The minimal number of inflationary $e$-folds required in order for the associated no-boundary instantons to be allowable given the potential \eqref{eq:potential}. The quadratic fit to the numerical solutions is given by $N_\mathrm{min,allow}=1.720-3.831f+2.257f^2$.}
	\label{fig:f_vs_N_min_allow}
\end{figure}

We can illustrate this with an explicit example, for the family of potentials given by the (natural inflation \cite{Freese:1990rb}) form 
\begin{equation}
    V(\phi) = V_0 \left( 1 + \cos(\phi/f)\right)\,,\label{eq:potential}
\end{equation}
where $f$ is a parameter. Then we find that small amounts of inflation lead to non-allowable instantons and are thus excluded. But above a minimal amount $N_\mathrm{min,allow}$ the instanton has sufficiently small complex parts of the metric to be allowed. The specific minimal number of $e$-folds depends on the parameter $f$, as shown in Fig.~\ref{fig:f_vs_N_min_allow}. For example, when $f=4$ we find $N_\mathrm{min,allow}\approx 22.5$, while for $f=6$ we find $N_\mathrm{min,allow} \approx 59.5$.

\begin{figure}[ht]
	\centering
	\includegraphics[width=0.4\textwidth]{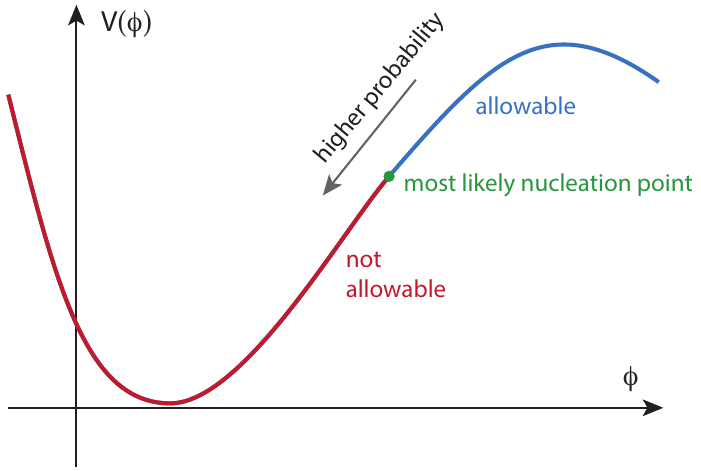}
	\caption{The scalar inflationary potential. The no-boundary wave function assigns higher probability to starting lower on the potential, but if one start too low, the associated instanton is not allowable. Thus the most likely history is the one that has the minimal number of $e$-folds required to make it allowable.}
	\label{fig:potential}
\end{figure}

If we combine this finding with the preference of no-boundary probabilities for starting out low on the potential, we find that the most likely histories are those that have precisely the minimal number of $e$-folds required for allowability. This situation is sketched in Fig.~\ref{fig:potential}.

\begin{figure}[ht]
	\centering
	\includegraphics[width=0.47\textwidth]{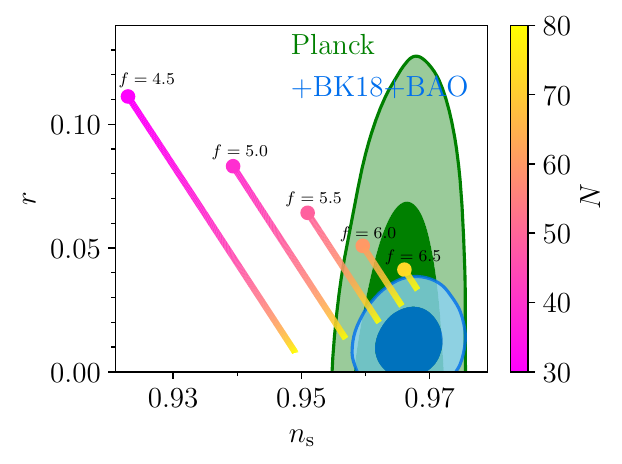}
	\caption{A plot of the predictions of the natural inflation models considered here with various parameter values $f$ and compared to the observational bounds obtained by the {\it Planck} satellite \cite{Planck:2018jri} (in green) and the BICEP/{\it Keck} experiments \cite{BICEP:2021xfz} (in blue). The colour coding depicts the number of $e$-folds from the time of horizon exit until the end of inflation; $N=N_\mathrm{min,allow}$ corresponds to the large dots, as per our results shown in Fig.~\ref{fig:f_vs_N_min_allow}.}
	\label{fig:rns}
\end{figure}

\begin{figure}[ht]
	\centering
	\includegraphics[width=0.47\textwidth]{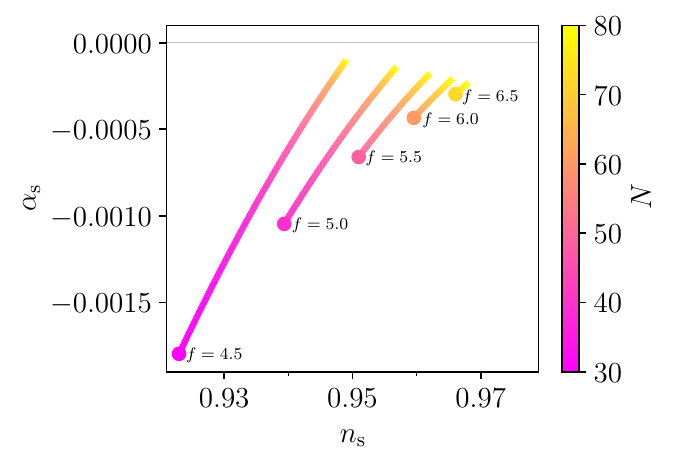}
	\caption{For the same models, predictions for the running of the spectral index $\alpha_\mathrm{s} = \dd n_\mathrm{s}/\dd N$.}
	\label{fig:alphans}
\end{figure}

We should now see whether these preferred solutions can also be in agreement with observations. Their predictions for the spectral index $n_\mathrm{s}$ of density perturbations, as well as the amplitude of gravitational waves (quantified by the tensor-to-scalar ratio $r$), are shown in Fig.~\ref{fig:rns}, while the predictions for the running of the spectral index $\alpha_\mathrm{s}$ are in Fig.~\ref{fig:alphans}. Figure \ref{fig:rns} shows that the models that are in good agreement with observational data are those with $f \gtrapprox 6$. These models predict a fairly small running, within current observational bounds, but at a level that is detectable by not-too-distant future experiments.

A surprising aspect of this analysis is that {\it a priori} we have no idea what $N_\mathrm{min,allow}$ might be. It could have turned out to be $\mathcal{O}(1)$ or $\mathcal{O}(10^9)$ (and for some potentials it might very well reach such values), but it turned out to be of ${\cal O}(10)$ for models that are also in close agreement with observations. This is a remarkable feature. But we must also check whether these models actually solve the flatness puzzle. To this end, we must match the scale $V_0$ of the potential \eqref{eq:potential} with the observed amplitude of density perturbations. At the pivot scale we have \cite{Planck:2018jri} $\Delta^2\approx V_0/(8\pi^2\epsilon)\approx 2\times 10^{-9}$, which, given $\epsilon \sim 10^{-3}$, implies $V_0 \sim 10^{-10}$. This in turn implies a high reheating temperature of about $T_\mathrm{reh} \sim 10^{15}\,\textrm{GeV}$ and thus requires at least $60$ $e$-folds of inflation in order to explain the current large-scale flatness of the universe [{\it cf.}~\eqref{eq:Ninflmin}]. This is precisely what we obtain for models with $f \gtrapprox 6.$ Thus there is a kind of positive conspiracy at play, in that larger $f$ means more inflation and better agreement with observational data. It is remarkable that models of this kind provide a possible explanation for the otherwise rather arbitrary seeming value of $60$ inflationary $e$-folds.

Note that the $f \gtrapprox 6$ models predict a number of $e$-folds that is not much larger than the minimal amount required for obtaining a sufficiently flat universe. Thus, they also imply that the size of the universe is not much larger than the current observable size, illustrating the theme of this paper.

We emphasize that similar results are also obtained for other concave potentials that are in agreement with observational constraints on $\{n_\mathrm{s},r\}$ at the $2\sigma$ confidence level, bearing in mind that the theoretical predictions and the observed central values and error bars may shift when allowing for spatial curvature ({\it e.g.}~\cite{Handley:2019anl,Hergt:2022fxk,Letey:2022hdp,Park:2017xbl,Planck:2018vyg,Planck:2018jri,Handley:2019tkm}).

Now that we have seen with the example of an explicit model how allowable no-boundary instantons may imply a relatively small size of the universe, let us comment on a few consequences of this idea.

\section{Discussion}

We will start by asking how large we may expect $|\Omega_{K,0}|$ to be at present times. This depends very sensitively on the number of $e$-folds of inflation. The minimal number was determined precisely such that $|\Omega_{K,0}|=0.01$. However, $|\Omega_{K}| \propto e^{-2N_\mathrm{infl}}$ and thus if inflation lasts only $4$ $e$-folds longer than the minimal number, then $|\Omega_{K,0}| \sim 10^{-5}$. This is the lowest value we can theoretically detect, as the primordial density perturbations also induce perturbations on that scale (and it may in fact be difficult to obtain conclusive evidence below $|\Omega_{K,0}| \sim 10^{-4}$ \cite{Vardanyan:2009ft,Leonard:2016evk}). Thus it still seems possible to directly detect whether or not our universe possesses a small size, but only if inflation lasted little more than the minimal required duration.
 
Conceptually, a small universe seems to fit rather well with the recent advances in string theory and in particular the restrictions on solutions codified in the swampland programme. These suggest that inflation is hard to obtain: the dS conjecture \cite{Obied:2018sgi,Garg:2018reu,Ooguri:2018wrx,Lehners:2018vgi} implies that either $\epsilon$ or $\eta\equiv V_{,\phi\phi}/V$ must be large. Plateau models (with small $\epsilon$ and large $\eta$) are particularly well suited: if you start higher up the potential, where it is flatter, then instantons will be more real and more allowable, as illustrated with the examples above. If the trans-Planckian censorship conjecture holds \cite{Bedroya:2019snp,Bedroya:2019tba,Kadota:2019dol}, then it would imply a very small $\epsilon$, and thus also a very low inflationary scale (perhaps as low as $1\,\textrm{TeV}$). This would require a very modest number of $e$-folds of inflation, corresponding to a low reheating temperature and thus a smaller required number of $e$-folds in order to resolve flatness and horizon puzzles. Moreover, a small $\epsilon$ also makes no-boundary geometries almost real [{\it cf.}~\eqref{eq:phiSP}]. Specific predictions however can only be made in the presence of concrete constructions of potentials -- we will thus leave this question for future work.

Another consequence of the swampland conjectures is that they must also apply to the current dark energy phase. Then they imply that dark energy must decay. If its potential becomes negative, the universe may even recollapse in a short period of time (see \cite{Andrei:2022rhi} for a study of this scenario). This would then imply a universe that is not only small spatially, but also in temporal extent. This would have the further consequence of rendering the universe rather stable to vacuum decay.

A further idea that has emerged from the swampland programme is that of an additional mesoscopic spatial dimension, called the dark dimension (see, {\it e.g.}, \cite{Montero:2022prj,Anchordoqui:2022svl} and references therein). One puzzle with this scenario is how this dimension can be combined with inflation occurring in the other large dimensions, due to the Higuchi bound which effectively implies that there can be no extra internal dimension that is larger than the size of an inflating universe \cite{Higuchi:1986py}. Here we see a possible resolution: the universe can simply nucleate at a large size already, corresponding to a low inflationary scale, in an isotropic $5$-dimensional configuration. One might then imagine a mechanism by which $3$ spatial dimensions inflate, while one dimension is held fixed.

As a concluding thought, let us point out that dark energy currently provides the largest physical scale known. Understanding it will likely require a better understanding of quantum gravity, in perfect illustration of the emerging principle that quantum gravity affects both the smallest and largest scales accessible to us. Here we have seen that quantum cosmology gives us a reason to think that the size of the entire universe may be given by the same scale and may not be more than a few orders of magnitude larger than the currently visible part.

\section*{Acknowledgements}
JLL thanks Oliver Janssen for explaining the numerical methods used in \cite{Hertog:2023vot}.
JLL gratefully acknowledges the support of the European Research Council via the ERC Consolidator Grant CoG 772295 ``Qosmology''.
This research was supported in part by the Perimeter Institute for Theoretical Physics. Research at the Perimeter Institute is supported by the Government of Canada through the Department of Innovation, Science and Economic Development and by the Province of Ontario through the Ministry of Colleges and Universities.

\appendix

\section{Allowable complex paths}\label{app:details}

We consider general relativity minimally coupled to a scalar field with a potential $V(\phi);$ in natural units ($8\pi G=c=1$) and with the cosmological metric
\begin{equation}
    g_{\alpha\beta}\dd x^\alpha\dd x^\beta=\dd\tau^2+a(\tau)^2\dd\Omega_{(3)}^2\,,
\end{equation}
the equations of motion read
\begin{subequations}
\begin{align}
    & a_{,\tau\tau}+\frac{a}{3}\left((\phi_{,\tau})^{2}+V(\phi)\right)=0\,,\\
    & \phi_{,\tau\tau}+3\frac{a_{,\tau}}{a}\phi_{,\tau}-V_{,\phi}=0\,,
\end{align}
\end{subequations}
while the constraint is 
\begin{equation}
    (a_{,\tau})^2-1=\frac{a^2}{3}\left( \frac{1}{2}(\phi_{,\tau})^2 - V(\phi)\right)\,.
\end{equation}
No-boundary solutions are then regular solutions satisfying $a_{,\tau}=1$ at the South Pole $\tau=0$ and reaching specified final values $a(\tau_\mathrm{f})=b$, $\phi(\tau_\mathrm{f})=\chi$ at a final time $\tau_\mathrm{f}$ (see, {\it e.g.}, \cite{Lehners:2023yrj}). In order to obtain such solutions one must find a suitable (complex) scalar field value $\phi_\mathrm{SP}$ at $\tau=0$. This works (typically with the use of a numerical optimisation algorithm) as long as the potential $V(\phi)$ is such that it leads to a dynamical attractor, in our case inflation. In the slow-roll case one typically has
\begin{equation}
    \Re(\phi_\mathrm{SP}) \approx \chi\,, \qquad \Re(\tau_\mathrm{f})\approx \frac{1}{2\pi}\sqrt{\frac{V(\chi)}{3}}\,,
\end{equation}
while
\begin{align}
    \Im(\phi_\mathrm{SP})&\approx - \frac{\pi}{2}\frac{V_{,\phi}}{V}\,,\nonumber\\
    \qquad \Im(\tau_\mathrm{f})&\approx\sqrt{\frac{3}{V(\chi)}}\,\mathrm{arcosh}\left(b\sqrt{\frac{V(\chi)}{3}}\right)\,.
\end{align}
One may then think of the no-boundary solution as a path in the complex time plane starting at $\tau=0$ and ending at $\tau_\mathrm{f}$; see Fig.~\ref{fig:path}. Reference \cite{Janssen:2020pii} provides further details.

We would now like to know whether the solution along this path (let us call it $\gamma(u),$ with real parameter $u$) is allowable or not \cite{Jonas:2022uqb,Hertog:2023vot}. For this, we may directly solve the equations of motion along the path. The metric then becomes
\begin{equation}
    \dd s^2=\gamma^{\prime 2}\dd u^2+a^2 \dd\Omega_{(3)}^2\,,
\end{equation}
where a prime denotes a derivatives with respect to $u$. Note that $\gamma$ is generally complex valued.  Then using the chain rule ($\dd a/\dd\tau=a'/\gamma'$) we get the equations of motion along the path,
\begin{subequations}
\begin{align}
    & a''-\frac{\gamma''}{\gamma'}a'+\frac{a}{3}\left(\phi^{\prime 2}+\gamma^{\prime 2}V(\phi)\right)=0\,,\\
    & \phi''-\frac{\gamma''}{\gamma'}\phi'+3\frac{a'}{a}\phi'-\gamma^{\prime 2}V_{,\phi}=0\,.
\end{align}
\end{subequations}
The equation for the maximal allowed curve is [{\it cf.}~saturating the inequality \eqref{bound}]
\begin{equation}
    \left|\mathrm{Arg}\left(\gamma^{\prime 2}\right)\right|+3\left|\mathrm{Arg}(a^2)\right|=\pi\,,
\end{equation}
where for the maximal curve we can remove the absolute value signs (the curve always moves up and to the right in the region of interest, and the imaginary part of $a$ is positive and small for the solutions of interest compared to the positive real part). Only the argument of $\gamma'$ is fixed, and a solution with unit modulus is
\begin{equation}
    \gamma'=i\left(\frac{a^*}{a}\right)^{3/2}\,.
\end{equation}
We can also take a derivative to obtain an expression for $\gamma''$,
\begin{equation}
    \frac{\gamma''}{\gamma'}=\frac{3}{2}\left(\frac{a^{*\prime}}{a^*}-\frac{a'}{a}\right)\,,
\end{equation}
which can then be substituted into the equations of motion, ending up with
\begin{subequations}
\begin{align}
    &a''-\frac{3}{2}\left(\frac{a^{*\prime}}{a^*}-\frac{a'}{a}\right)a'+\frac{a}{3}\left(\phi^{\prime 2}-\left(\frac{a^*}{a}\right)^{3}V(\phi)\right)=0\,,\\
    &\phi''-\frac{3}{2}\left(\frac{a^{*\prime}}{a^*}-\frac{a'}{a}\right)\phi'+3\frac{a'}{a}\phi'+\left(\frac{a^*}{a}\right)^{3}V_{,\phi}=0\,.
\end{align}
\end{subequations}
These are the equations that are then solved numerically in order to obtain the results shown in Fig.~\ref{fig:f_vs_N_min_allow}.  We find that the level of numerical precision that is required is very high. For instance, we find that for solutions expanding by $N$ $e$-folds, we typically need to retain about $3N$ decimal places in the numerical calculations. As described in the main text, if $\tau_\mathrm{f}$ lies below the maximal curve, then an allowable path can be found. Otherwise we deem the solution unphysical.

\bibliographystyle{JHEP2}
\bibliography{refs}

\end{document}